\documentclass[12pt]{article} \input epsf.tex

\newcommand{\be}{\begin{equation}}
\newcommand{\ee}{\end{equation}}
\newcommand{\bea}{\begin{eqnarray}}
\newcommand{\eea}{\end{eqnarray}} 
\newcommand{\ba}{\begin{array}}
\newcommand{\ea}{\end{array}}

\begin{document}

\begin{titlepage}
\def\thepage {}        

\title{Neutrino mixing in the seesaw model}

\author{
T. K. Kuo$^1$, Shao-Hsuan Chiu$^2$, Guo-Hong Wu$^3$ \\ \\ 
{\small {\it
$^1$Department of Physics, Purdue University, West Lafayette, IN 47907
\thanks{Email: tkkuo@physics.purdue.edu, sxcsps@rit.edu, 
gwu@darkwing.uoregon.edu}}}\\
{\small {\it $^2$Department of Physics, Rochester Institute of Technology,
         Rochester, NY 14623}}\\
{\small  {\it and}}\\
{\small {\it  Department of Physics, Frostburg State University
         Frostburg, MD 21532}}\\
{\small {\it $^3$Institute of Theoretical Science, University of Oregon,
Eugene, OR 97403}}\\ 
}

\date{Revised, May 2001}   \maketitle

\begin{abstract}
In the seesaw model with hierarchical Dirac masses, the neutrino mixing
angle exhibits the behavior of a narrow resonance.  In general,
the angle is strongly suppressed, but it can be maximal for
special parameter values.  We delineate the small regions in which
this happens, for the two flavor problem.  On the other hand, the physical
neutrino masses are hierarchical, in general, 
except in a large part of the region in
which the mixing angle is sizable, where they are nearly degenerate.
 Our general analysis is also applicable to the RGE of neutrino
mass matrix, where we find analytic solutions for the running of
physical parameters, in addition to a complex RGE invariant relating
them.
It is also shown that, if one mixing angle is small, the three neutrino
problem reduces to two, two flavor problems.

\end{abstract}


\vfill \end{titlepage}

\section{Introduction}

	The exciting development of recent experiments~\cite{sk} has offered strong
evidence for the existence of neutrino oscillation, from which one can infer
about the intrinsic properties of the neutrinos.  While the neutrino
masses (mass differences) are found to be very tiny, there is a major
surprise for the mixing angles.  It is found that at least one, and
possibly two, of the three mixing angles are large, or even maximal.
This is in stark contrast to the situation in the quark sector, where
all mixing angles are small.

	Theoretically, the seesaw model~\cite{gellmann} is very appealing in that it
can offer a natural mechanism which yields small neutrino masses.
However, owing to its complex matrix structure, it is not obvious what
the implied patterns of neutrino mixing are.  In a previous paper~\cite{kuo}, we
found a parametrization which enabled us to obtain an exact solution
to the two flavor seesaw model.  When one makes the usual assumption 
that the Dirac mass matrix has a strong hierarchy, 
the physical neutrino mixing angle exhibits the narrow resonance behavior.
For generic parameters in the Majorana mass matrix, the physical
neutrino mixing angle is strongly suppressed.  However, if the parameters
happen to lie in a very narrow region, the mixing can be maximal.

	In this paper we will expand on our earlier investigations
and discuss in detail the behavior of the neutrino mixing matrix in
the seesaw model.  As was shown before, if we assume that the Dirac mass
hierarchy is similar to that of the quarks, the problem has three relevant
parameters associated with the Majorana sector,
 namely, the mixing angle, the ratio of masses,
and their relative phase.  We will present
plots of the physical neutrino mixing angle and their mass ratio in the
3D parameter space.  These will offer a bird's-eye view of their behaviors.
In particular, the neutrino mixing angle is only appreciable in a very small region, 
which we exhibit explicitly.  Furthermore, this region is complementary to the 
region in which there is appreciable physical neutrino mass hierarchy.
Thus, roughly speaking, the seesaw model divides the 3D parameter space in two parts.
There is a very small region in which the mixing angle is large, at the same
time the neutrino masses are nearly degenerate.  For most parameters, the
mixing angle is small but there is a strong hierarchy in the mass eigenvalues.
An exception to this picture is when the Majorana matrix has extreme
hierarchy and very small mixing angle.  In this tiny region, the
physical neutrinos can be hierarchical and simultaneously their mixing
angle is large.

	The solution to the seesaw problem is most transparent in the 
parametrization introduced in Ref.~\cite{kuo}.  However, it is useful to make 
connections with the usual notation, where individual matrix elements
are regarded as independent parameters.  
We obtain relations which clarify the roles played by the
various parameters.  They enable one to gain insights in understanding
the numerical results presented in the 3D plots.

   The general analysis of symmetric and complex matrices turns out
to be useful in other applications. Our method can be used to yield
an analytic solution of the renormalization group equation (RGE) of the
effective neutrino mass matrix. In addition, we obtain a (complex) RGE
invariant which relates the running of the mixing angle and the complex mass
ratio. The detailed analysis of the structure of the seesaw matrix also
suggests a universal picture for the quark as well as the neutrino mass 
matrices. While the quarks have generally small mixing angles and 
hierarchical mass ratios, if one assumes that the Majorana matrix itself 
is of the seesaw type, the effective neutrino mixing angle can be
naturally large.

	Finally, we turn to a discussion of the three neutrino problem.
Although the principle involved here is the same as in the two neutrino problem,
the algebra with the Gell-Mann $\lambda$ matrices is far more complicated than
that of the Pauli $\sigma$ matrices.  We are unable to obtain a general solution
in this case.  However, it is quite well-established that one of the neutrino
mixing angles is small~\cite{chooz}.  
In this case, an approximate solution can be obtained.
It turns out that, to lowest order, the three neutrino problem can be 
reduced to two, two-flavor problems.  This solution can thus accommodate the
``single-maximal" or ``bimaximal" solutions that have been considered in
the literature.


\section{The two flavor problem}
\label{sec:2flav}

	In a previous paper~\cite{kuo,st}, an exact solution was obtained for the two 
flavor seesaw model.  In this section, in addition to a summary of 
the earlier paper, further results will be presented.

	For two flavors, the seesaw model 
\begin{equation} \label{eq:seesaw} 
 		m_{\nu}=m_{D}M_{R}^{-1}m_{D}^{T},
\end{equation}
can be written in the form
\begin{equation}  \label{eq:MN}
	m_{\nu}=U \left(\begin{array}{cc}
                m_{1} &  \\
                  & m_{2} \\
                 \end{array}
                  \right) V_{R}
       \left(\begin{array}{cc}
                M_{1}^{-1} &  \\
                  & M_{2}^{-1} \\
                 \end{array}
                  \right) V_{R}^{T}            
        \left(\begin{array}{cc}
                m_{1} &  \\
                  & m_{2} \\
                 \end{array}
                  \right) U^{T}
\end{equation}
Let us introduce the parametrization
\begin{equation}
	\left(\begin{array}{cc}
                m_{1} &  \\
                  & m_{2} \\
                 \end{array}
                  \right)=\sqrt{m_{1}m_{2}} e^{-\xi \sigma_{3}},
   \mbox{  } \,\,\,\, \xi  =  \frac{1}{2}\ln(m_{2}/m_{1});
\end{equation}
\begin{equation}
 	\left(\begin{array}{cc}
                M_{1}^{-1} &  \\
                  & M_{2}^{-1} \\
                 \end{array}
                  \right)=\sqrt{\frac{1}{M_{1}M_{2}}} e^{2\eta \sigma_{3}},
           \mbox{  } \,\,\,\,   \eta  =  \frac{1}{4}\ln(M_{2}/M_{1});
\end{equation}
\begin{equation} \label{eq:V}
	V_{R}=e^{i\alpha \sigma_{3}}e^{-i\beta \sigma_{2}}
	e^{i\gamma \sigma_{3}}.
\end{equation}
Thus, in the basis in which $m_{D}$ is diagonal, $\beta$ is the mixing
angle for $M_{R}^{-1}$ while $\pm 2\gamma$ are the phases of the eigenvalues.
This parametrization shows clearly that the relevant variables in the
diagonalization of $m_{\nu}$ are $\xi$, $\alpha$, $\beta$, $\gamma$ and
$\eta$.  Of these, it is usually assumed that $m_{2}/m_{1}$ can be identified
with the known quark mass ratio.  Also, $\alpha$ can be absorbed into
$U$ as part of the phase of the Dirac mass eigenvalues.  For $U \simeq I$,
in particular, it becomes the phase of the charged leptons and is not
observable.

	Note also that, apart from an overall constant, $m_{\nu}$ is a
product of $2 \times 2$, complex matrices with $det = +1$, i.e., it is
an element of $SL(2,C)$.  Thus, we can identify $m_{\nu}$ with an
element of the Lorentz group, with $\xi$ and $\eta$ interpreted as
rapidity variables.

	To find the effective neutrino mixing matrix, we need to 
rearrange the matrices in $m_{\nu}$ in a different order
\begin{equation} \label{eq:mW}
	m_{\nu}=\sqrt{\frac{m_{1}^{2}m_{2}^{2}}{M_{1}M_{2}}}UW
                e^{-2\lambda \sigma_{3}}W^{T}U^{T},
\end{equation}

\begin{equation}\label{eq:W} 
	W=e^{i\omega' \sigma_{3}}e^{-i\theta \sigma_{2}}e^{i\phi \sigma_{3}},
\end{equation}
\begin{equation}
	\omega' \equiv \omega + \alpha,  \mbox{   } \,\,\,\, 
	\lambda = \frac{1}{4} \ln(\mu_{2}/\mu_{1}).
\end{equation}
Here, the physical neutrino masses are given by $\mu_{1}$ and $\mu_{2}$,
with their ratio given in terms of $\lambda$, while $4\phi$ is their
relative phase.  We have also absorbed the phase $\alpha$ into $\omega'$.
The physical neutrino mixing matrix is given by $UW$, so that $W$ is the
induced mixing matrix from the seesaw mechanism.  The left-handed Dirac mixing,
$U$, in analogy to the quark sector, is often taken to be close 
to the identity, $U \simeq I$.  In the following we will concentrate on the
behavior of $W$ only, corresponding to $U \simeq I$.  
However, when necessary, $U$ can always be included in the final result.

	As was shown before, the solution for $W$ corresponds to that of
the velocity addition problem in relativity,  and one can readily obtain
the answer by manipulating the Pauli matrices.  We have~\cite{kwc} 
\begin{equation} \label{eq:tomega}
  \tan 2\omega=\frac{\Sigma_{I}}{
      \Sigma_{R}\coth 2\xi-\cos 2\beta},
\end{equation}
\begin{equation} \label{eq:ttheta}
	 \tan 2\theta=
	\frac{\sin 2\beta/(\cos 2\omega \cosh 2\xi)}{\cos 2\beta-\Sigma_{R}
 	\tanh 2\xi-\Sigma_{I}\tan 2\omega},
\end{equation}
\begin{equation} \label{eq:clambda}
 \cosh 2\overline{\lambda}=
\cosh 2\bar{\xi} \cosh 2\bar{\eta}-\cos 2\beta \sinh 2\bar{\xi}
        \sinh 2\bar{\eta},
\end{equation}
where $\bar{\lambda}=\lambda+i\phi$, $\bar{\xi}=\xi - i\omega$,
$\bar{\eta}=\eta + i\gamma$, and
\begin{eqnarray} \label{eq:ceta}
\coth 2\overline{\eta} & = & \frac{1-(M_{1}/M_{2})^{2}-2i(M_{1}/M_{2})\sin 4\gamma}
   {1+(M_{1}/M_{2})^{2}-2(M_{1}/M_{2})\cos 4\gamma} \nonumber  \\
  &  \equiv & \Sigma_{R}+i\Sigma_{I}.
\end{eqnarray}

	Note the non-trivial contribution from $\tan 2\omega$ in 
Eq.~(\ref{eq:ttheta}).  To diagonalize the symmetric and complex
mass matrix, $U^{-1}m_{\nu}U^{*}$, as is detailed in the next section,
it is necessary to multiply the mass matrix on either side by the same
phase matrix.  This phase matrix is precisely $e^{-i\omega' \sigma_{3}}$.

	Eq.~(\ref{eq:ttheta}) shows that, when $m_{D}$ is hierarchical
($\xi \gg 1$), the neutrino mixing angle $\theta$ is small
($\tan \theta \sim e^{-2\xi} \sim m_{1}/m_{2}$), for
generic values of the other parameters, $\beta$, $\eta$, and $\gamma$.
However, when the denominator in Eq.~(\ref{eq:ttheta}) vanishes,
$\theta$ is maximal.  This is the resonance behavior mentioned before.
In general, the seesaw mechanism suppresses the neutrino mixing angle.
But when the resonance condition is met, it is enhanced and becomes maximal.

	This behavior is quantified in Fig.1, which is a 3D plot of the
region $\sin^2 2\theta > 0.5$, within the parameter space spanned by
$\cos 2\beta$, $\gamma$ and $M_{1}/M_{2}$.  This region consists
roughly of two parts.  One runs along the edge $\cos 2\beta \approx 1$
and $M_{1}/M_{2} \ll 1$, but $\gamma$ can take values between 0 and
$\pi/4$.  The other region is tube-like, and ``hugs" the back wall,
$\gamma \approx \pi/4$, with $\cos 2\beta \approx \tanh 2 \eta$.
It is striking how small the region for $\sin^2 2\theta >0.5$ is.
Outside of this region, which consists of most of the parameter space,
$\sin^2 2\theta$ is tiny ($\sim (m_{1}/m_{2})^2$).  This result is
the analog of the familiar focusing mechanism in relativity.
When a relativistic particle decays, most of the decay products are
contained in a forward cone of opening angle $\leq 1/\gamma_{0}$, where
$\gamma_{0} = 1/\sqrt{1-v^{2}/c^{2}}$.  This corresponds to the seesaw
problem with the identification 
$\gamma_{0}=\cosh 2\xi \simeq \frac{1}{2}(m_{2}/m_{1})$.

	In Fig.2, we blow up the region with a fixed 
$\cos 2\beta \approx 1$.  It is seen that there is considerable structure
when $\sin^2 2\theta$ is maximal.  In particular, the dependence on
$\gamma$ is highly non-trivial.  From the scale in the figure, we see
that large values of $\sin^2 2\theta$ are confined in a very narrow region
with width $\sim (m_{1}/m_{2})^2$.
Note also that, outside of the maximal $\sin^{2}2\theta$ region, near the
upper left edge of Fig.1 ($M_{1}/M_{2} \rightarrow 0, \cos 2\beta \rightarrow 1$),
$\sin^{2} 2\theta$ remains large.  This region
is characterized by extreme hierarchy of the Majorana masses
$((m_{1}/m_{2})^{2}>(M_{1}/M_{2}) \rightarrow 0)$ and very small
$\beta$ ($(1-\cos 2\beta) \sim (m_{1}/m_{2})^{2}$).

	Fig.3 shows the contents of Fig.1 in a 2D parameter space,
with $\cos 2\beta = 1/2$.  It exhibits clearly the behavior of
$\sin^2 2\theta$ near $\gamma=\pi/4$.  Here, the maximum of $\sin^2 2\theta$
is attained at $\cos 2\beta =\tanh 2\eta \cdot \tanh 2\xi$ with $\gamma=\pi/4$.
Away from these values, $\sin^2 2\theta$ drops off quickly.  The width of
the peak is of order $(m_{1}/m_{2})$ in either $\Delta(M_{1}/M_{2})$ or
$\Delta \gamma$.

	The behavior of physical neutrino mass ratio is depicted in
Fig.4, which exhibits the region of near degeneracy, $\mu_{1}/\mu_{2} >0.5$.
We have chosen a log scale for $M_{1}/M_{2}$ to highlight the detailed
structure near the upper left edge.  A comparison with Figs. 1 and 2 reveals the
complementary nature of the regions of maximal $\sin^{2} 2\theta$ 
versus hierarchical $\mu_{1}/\mu_{2}$.
In the small region where $\sin^2 2\theta \simeq 1$, one also has
$\mu_{1}/\mu_{2} \simeq 1$.  However, near the upper left
edge, for $(m_{1}/m_{2})^{2}> M_{1}/M_{2} \rightarrow 0$, 
$\mu_{1}/\mu_{2}$ can be small and at the same time $\sin^2 2\theta$ is large.  For generic parameters,
$\sin^2 2\theta \sim (m_{1}/m_{2})^2$, but the masses are also very hierarchical,
$(\mu_{1}/\mu_{2}) \sim (m_{1}/m_{2})^{2}$.

\section{General properties of mass matrices}
\label{sec:GP}

	In order to gain some insights about the results presented in the
previous section, it is useful to study the general properties of symmetric,
complex, matrices.  We will first discuss the relations between different
parametrizations of the mass matrices.  These relations will shed light
on the special properties of matrices of the seesaw type.  They will also
enable one to have a qualitative understanding of the results presented in
Sec.~\ref{sec:2flav}.

\subsection{Parametrization of neutrino mass matrices}

	Within the framework of the seesaw model, the neutrinos are 
Majorana in nature, so that their matrices are symmetric and complex, 
in general.  We first consider the case of two flavors,
\begin{equation}  \label{eq:N0}
	N=\left(\begin{array}{cc}
                A & B \\
                B  & C \\
                 \end{array}
                  \right).
\end{equation}
Here, $A, B$, and $C$ are arbitrary complex numbers.  Without loss of
generality, we assume that $N$ is normalized so that
$det N$ = +1,
\begin{equation}
	AC-B^{2}=1.
\end{equation}
This matrix can be diagonalized by a unitary matrix $U$,
\begin{equation}
	N=U e^{2\eta \sigma_{3}} U^{T}.
\end{equation}
In terms of the eigenvalues ($n_{1},n_{2}$), 
$\eta=\frac{1}{4} \ln(n_{2}/n_{1})$.  A convenient choice for
$U$ is in the Euler parametrization
\begin{equation}
	U=e^{i\alpha \sigma_{3}}e^{-i\beta \sigma_{2}}e^{i\gamma \sigma_{3}}.
\end{equation} 
The relation between the two parametrizations of $N$ is given by
\begin{equation} \label{eq:N}
	\left(\begin{array}{cc}
                A & B \\
                B  & C \\
                 \end{array}
                  \right)=
\left(\begin{array}{cc}
                e^{2i\alpha}(\mathrm{ch}2\bar{\eta} +
                     \mathrm{C}_{2\beta} \cdot sh2\bar{\eta}) 
                & \mathrm{S}_{2\beta}\cdot \mathrm{sh}2\bar{\eta}   \\
                \mathrm{S}_{2\beta}\cdot \mathrm{sh}2\bar{\eta}  & 
                e^{-2i\alpha}(\mathrm{ch}2\bar{\eta}- 
               \mathrm{C}_{2\beta} \cdot \mathrm{sh}2\bar{\eta})    \\
                 \end{array}
                  \right),
\end{equation}
where we have used the notation $\bar{\eta}=\eta+i\gamma$, 
$\mathrm{ch}2\eta \equiv \cosh 2\eta$,
$\mathrm{S}_{2\beta} \equiv \sin 2\beta$, etc.

	Note that because of the condition $AC-B^{2}=1$, there are exactly
four parameters in the three complex numbers $A, B$, and $C$.  To understand
the role played by the phase $\alpha$, let us write
\begin{equation}\label{eq:abc}
	N=\frac{1}{2}(A+C)+\frac{1}{2}(A-C)\sigma_{3}+B\sigma_{1}.
\end{equation}
The diagonalization of $N$ is easy provided that the phase of
$A-C$ and $B$ are the same.  In general, we can multiply $N$
on either side by the same phase, $e^{-i\alpha \sigma_{3}}$,
\begin{equation}
	e^{-i\alpha \sigma_{3}} N e^{-i \alpha \sigma_{3}}=
	\frac{1}{2}(e^{-2i \alpha}A+e^{2i \alpha} C) +
	\frac{1}{2}(e^{-2i \alpha}A-e^{2i \alpha}C)\sigma_{3}
                   +B \sigma_{1}.
\end{equation}
We now choose $\alpha$ so that the phase of 
$(e^{-2i \alpha}A-e^{2i \alpha}C)$ coincides with that of $B$:
\begin{equation} \label{eq:alpha}
	\arg B=\arg (e^{-2i \alpha}A-e^{2i \alpha}C).
\end{equation}
In  this case, the matrix 
$e^{-i\alpha \sigma_{3}} N e^{-i\alpha \sigma_{3}}$ can be 
diagonalized by 
$e^{-i \beta \sigma_{2}}(e^{-i\alpha \sigma_{3}} 
N e^{-i\alpha \sigma_{3}})e^{i \beta \sigma_{2}}$, with
\begin{equation} \label{eq:beta}
	\tan 2\beta=\frac{2B}{e^{-2i \alpha}A-e^{2i \alpha}C}=real.
\end{equation}
Thus, given an arbitrary $N$, we need first to determine the
phase $\alpha$ by Eq.~(\ref{eq:alpha}).  After which $\beta$ is fixed by
Eq.~(\ref{eq:beta}), and then $\gamma$ can be read off from the 
diagonal matrix.  

Note that from Eq.(\ref{eq:beta}),
\be \label{eq:real}
B(e^{2i\alpha}  A^* - e^{-2i\alpha} C^* ) 
= B^* (e^{-2i\alpha}  A - e^{2i\alpha} C )   \ .
\ee
Thus, Eq.(\ref{eq:alpha}) is equivalent to 
\be \label{eq:2alpha}
2 \alpha = \arg ( AB^* + B C^*) \ .
\ee
In addition, we may use Eq.(\ref{eq:real}) in Eq.(\ref{eq:beta}) to 
obtain
\be \label{eq:2beta}
\tan 2\beta= {2 |AB^* + BC^*| \over |A|^2 -|C|^2} \ .
\ee
Note also that, if $|A|=|C|$, Eq.(\ref{eq:alpha}) can not be used to
solve for $\alpha$, but Eqs.(\ref{eq:2alpha}) and (\ref{eq:2beta})
are still valid. The complex eigenvalues of Eq.~(\ref{eq:N0}) can be
obtained from Eq.~(\ref{eq:N}). They are given by
\bea
 e^{+2\bar{\eta}} & = & A e^{-2 i \alpha}
                      + B \tan\beta  \\
 e^{-2\bar{\eta}} & = & C e^{+2 i \alpha}
                      - B \tan\beta   \ .
\eea

	It is also useful to introduce another variable,
\begin{equation}
	\bar{\zeta}=\frac{1}{2}\ln(A/C).
\end{equation} \label{eq:b1}
Using $\bar{\zeta}$, Eq.~(\ref{eq:beta}) can be written as
\begin{equation} \label{eq:betaprime} 
	\tan 2\beta=\frac{B/\sqrt{AC}}{\mathrm{sh}(\bar{\zeta}-2i\alpha)}.
\end{equation}
Also, from Eq.~(\ref{eq:N}),
\begin{eqnarray}
\mathrm{ch}2\bar{\eta} & = & \frac{1}{2}(e^{-2i\alpha}A+e^{2i\alpha}C) \nonumber \\                      & = & \sqrt{AC} \cdot \mathrm{ch}(\bar{\zeta}-2i\alpha).
\end{eqnarray}
Similarly,
\begin{equation}
	\mathrm{C}_{2\beta} \cdot \mathrm{sh}2\bar{\eta}=\sqrt{AC} 
   \cdot \mathrm{sh}(\bar{\zeta}-2i\alpha).
\end{equation}
We thus have
\begin{equation} \label{eq:zeta}
	\mathrm{C}_{2\beta} \cdot \tanh 2\bar{\eta}=\tanh(\bar{\zeta}-2i\alpha).
\end{equation}
This relation can be regarded as a consistency check on the properties
of $N$.  For instance, if $c_{2\beta}=0$ (maximal mixing),
it implies that Im$\bar{\zeta}=2\alpha$, and that Re$\bar{\zeta}=0$.
Another constraint is that the phase of $\tanh 2\bar{\eta}$ must be the
same as that of $\tanh (\bar{\zeta}-2i\alpha)$.

\subsection{The seesaw transformation}
\label{sec:ST}

	In the seesaw model, $m_{\nu}=m_{D}M_{R}^{-1}m_{D}^{T}$, it turns out
that the properties of $m_{\nu}$ is closely related to $M_{R}^{-1}$, when
we choose a basis in which $m_{D}$ is diagonal.  We shall call the change from
$M_{R}^{-1}$ to $m_{\nu}$ a ``seesaw transformation" (ST).  In terms of the 
notation of the previous section,  we define a ST from $N$ to a 
new matrix $M$ by
\begin{eqnarray}\label{eq:st} 
  M & =& e^{-\xi \sigma_{3}}Ne^{-\xi \sigma_{3}} \nonumber \\
   & = & \left(\begin{array}{cc}
                A' & B' \\
                 B' & C' \\
                 \end{array}
                  \right)
\end{eqnarray}
It is seen immediately that $B$ and $AC$ are invariant ($B'=B$, $A'C'=AC$), while
\begin{equation}
	A'/C'= e^{-4\xi}(A/C),
\end{equation}
or
\begin{equation}
	\bar{\zeta}'=\bar{\zeta}-2\xi,
\end{equation}
where $\bar{\zeta}'=\frac{1}{2}\ln(A'/C')$.
If we assume that 
\begin{equation}
	M=We^{2\lambda\sigma_{3}}W^{T},
\end{equation}
\begin{equation}
	W=e^{i(\omega + \alpha)\sigma_{3}}
	e^{-i\theta \sigma_{2}} e^{i\phi \sigma_{3}},
\end{equation}
we can use the results above to derive simple relations between the parameters
pertaining to $M$ and to $N$.  Thus, from the invariance
of $B$ under ST, we have immediately
\begin{equation} \label{eq:inv}
	\mathrm{S}_{2\beta} \cdot 
\mathrm{sh}2\bar{\eta}=\mathrm{S}_{2\theta} \cdot \mathrm{sh}2\bar{\lambda}, 
\end{equation}
i.e., $\mathrm{S}_{2\theta} \cdot \mathrm{sh}2\bar{\lambda}$ is an invariant, 
independent of $\xi$.
One of its consequences is that the phase of $\bar{\lambda}$ is tied to
that of $\bar{\eta}$, since $\beta$ and $\theta$ are both real.  In fact,
 if $\bar{\lambda}=\lambda + i\phi$, then
\begin{equation}
	\tan 2\phi \cdot \coth 2\lambda=
	\tan 2\gamma \cdot \coth 2\eta=constant,
\end{equation}
independent of $\xi$.  In particular, if $\bar{\eta}=\eta +i\pi/4$,
$\mathrm{sh} 2\bar{\eta}$ is purely imaginary, then the imaginary
part of $\lambda$ must also
be $\pi/4$, i.e., the mass eigenvalues must have opposite signs.
Moreover, given $\beta$ and $\bar{\eta}$, the relation exhibits 
the complementary nature of $\theta$ and $\lambda$, large $\theta$ 
correlates with small $\lambda$, and vice versa.  This behavior was already
discussed in connection with the results of Fig.4 in Sec.~\ref{sec:2flav}.
From Eq.~(\ref{eq:betaprime}), the invariance of $B/\sqrt{AC}$ yields
\begin{equation} \label{eq:theta}
	\tan 2\theta=\tan 2\beta \frac{\mathrm{sh}(\bar{\zeta}-2i\alpha)}
	{\mathrm{sh}(\bar{\zeta}-2\xi-2i\omega')}.
\end{equation}
When the ST is hierarchical, $\xi \gg 1$, it is clear that, for generic
$\bar{\zeta}$, the angle $\theta$ is suppressed 
$(\sim 1/\mathrm{sh}2\xi \sim (m_{1}/m_{2}))$.  However, if $\zeta \approx 2\xi$,
and if the phases in the denominator of Eq.~(\ref{eq:theta}) cancel, then
$\theta$ becomes maximal.  This was the behavior shown in Fig. 1.  It should
also be mentioned that Eq.~(\ref{eq:theta}) reduces to Eq.~(\ref{eq:ttheta})
when one uses Eq.~(\ref{eq:zeta}).

\begin{figure}[t]
\centerline{\epsfysize=20.cm\epsfbox{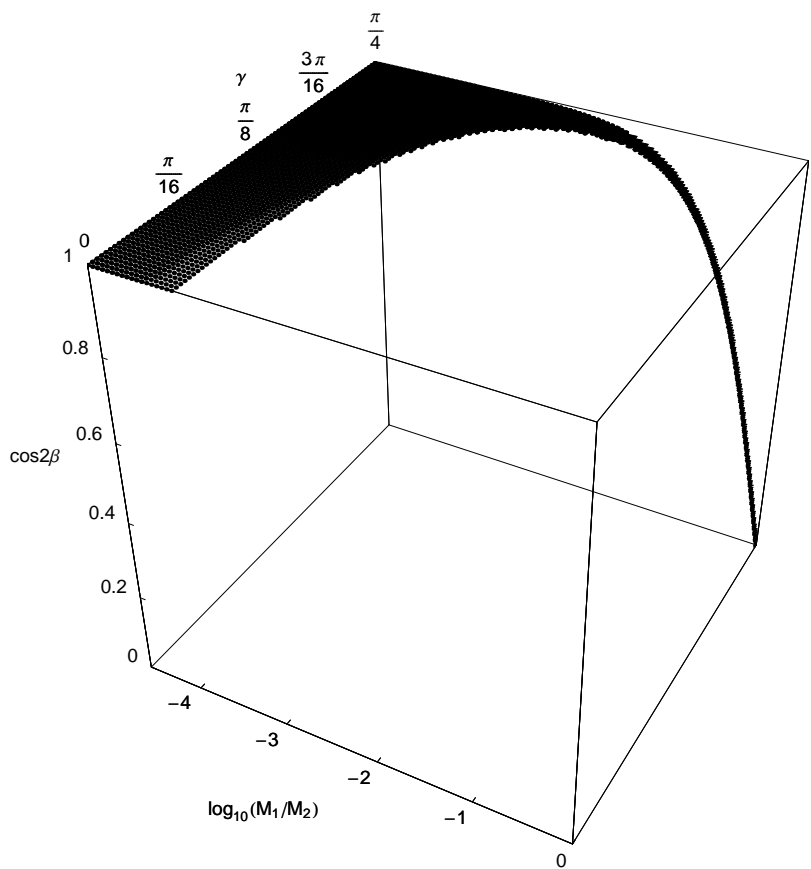}} 
\caption[] {Region in which $\sin^{2} 2\theta >0.5$, with
  $\tanh 2\xi = 0.9998$, or $(m_{1}/m_{2})=0.01$.  Note the log scale
used for ($M_{1}/M_{2}$).  \label{fig1.ps}} 
\end{figure}

	As another application, we note that a qualitative understanding of
Fig.1 can be gleaned from Eqs.~(\ref{eq:N}), ~(\ref{eq:beta}), ~(\ref{eq:st}).
Using Eqs.~(\ref{eq:beta}) and~(\ref{eq:st}), we have 
\be
\tan 2 \theta = {2B\over e^{-2i\omega^\prime} A^\prime 
- e^{2i\omega^\prime} C^\prime}
\ee
with $\omega=\omega^\prime -\alpha$ given by Eq.(\ref{eq:tomega}).
For $\xi \gg 1$, a necessary condition for large $\theta$ is that
$|C|\simeq 0$ (more precisely, $|C| \leq e^{-4\xi}|A|$ and 
$|C| \leq e^{-2\xi}|B|$). From Eq.(\ref{eq:N}), this means that
$\mathrm{ch} 2 \bar{\eta} \simeq \cos 2\beta \ \mathrm{sh} 2 \bar{\eta}$.
However, since $\mathrm{ch} 2 \bar{\eta}$ and $\mathrm{sh} 2 \bar{\eta}$
have different phases, this equation has only special solutions.
They are 1) $\gamma=\pi/4$, so that both 
$\mathrm{ch} 2 \bar{\eta}$ and $\mathrm{sh} 2 \bar{\eta}$
are purely imaginary, and $\cos 2 \beta \simeq \mathrm{coth}(2\eta+ i \pi/2)
=\mathrm{tanh} 2\eta$. This last equation describes the shaded trajectory
on the $\gamma=\pi/4$ wall in Fig.\ref{fig1.ps}.
Another solution is 2) $\eta \rightarrow \infty$, so that
$\mathrm{ch} 2 \bar{\eta} \simeq \mathrm{sh} 2 \bar{\eta} \simeq
e^{2i \gamma} e^{2\eta}/2$.
Then $\mathrm{ch} 2 \bar{\eta} - \cos 2\beta \ \mathrm{sh} 2 \bar{\eta}
\simeq e^{2i \gamma} e^{2\eta} (1-\cos 2\beta)/2$.
Since for $\eta \rightarrow \infty$, $|B|\simeq \sin 2\beta \  e^{2\eta}$,
$\theta$ can be large provided that $e^{2\xi}(1-\cos 2\beta) \ll \sin 2\beta$.
This solution corresponds to the shaded region in Fig.\ref{fig1.ps} with
$\gamma \neq \pi/4$.

  In summary, the neutrino mixing angle $\theta$ can only be large if the
$(2,2)$ element of $M_R^{-1}$ is small, $|C|\simeq 0$. The precise value
depends on phases and possible cancellation between $A^\prime$ and 
$C^\prime$. Note that, in the literature, a number of studies has concentrated
on the case of $M_R^{-1}$ being a real matrix.
For large mixing, two types of $M_R^{-1}$ have been identified.
1) $M_R^{-1} \simeq \left(\ba{cc} 1 & \epsilon \\ \epsilon & \epsilon^2 \ea
\right)$; 
2) $M_R^{-1} \simeq \left(\ba{cc} 0 & 1 \\ 1 & 0 \ea\right)$, or
$M_R^{-1} \simeq i \left(\ba{cc} 0 & 1 \\ 1 & 0 \ea\right)$, so that
$\mathrm{det} M_R^{-1}=+1$. 
These are special cases of $|C|\simeq 0$, corresponding to the two end points
of the shaded region in Fig.\ref{fig1.ps}, with coordinates 
$(M_1/M_2, \cos 2\beta, \gamma) \simeq (0,1,0)$ and $(1,0,\pi/4)$, 
respectively. Our analysis shows that care must be taken when we have 
small deviations from these forms, 
which arise naturally in models constructed from a presumed broken
symmetry. The narrowness of the shaded region means that a viable solution
can be easily thrown off course by small perturbations. An example of such
sensitivities is known in the renormalization group running effects, which we
will discuss in the next section.

\subsection{Renormalization}

  It turns out that our general analysis has an immediate application to the
renormalization group equation (RGE) analysis of the neutrino mass matrix.
We briefly comment on this connection. A full account will be given 
elsewhere\cite{kpw}.

  In the SM and MSSM, the RGE running of the neutrino mass matrix has been 
very extensively studied\cite{babu, RGE}. 
For simplicity, we only consider the two flavor problem 
with $(\nu_\mu, \nu_\tau)$. 
The RGE for the effective neutrino mass matrix is given by,
\be
{d\over dt} m_{\nu} = -(\kappa m_{\nu} + m_{\nu} P + P^T m_{\nu}) \ ,
\ee
where $\kappa$ is related to coupling constants, 
$t={1\over 16 \pi^2} \ln \mu/M_X$,
and to a good approximation,
\be
P \simeq P^T \simeq \chi(1-\sigma_3) \ ,
\ee
where $\chi$ is given by $y_\tau^2/4$ in the SM and $-\tilde{y}_\tau^2/2$ 
in the MSSM,
with $y_\tau$ and $\tilde{y}_\tau$ being the $\tau$ Yukawa coupling in
the SM and MSSM respectively.
The solution to RGE is 
\be
m_\nu(t) = e^{-\kappa^\prime t} e^{\xi \sigma_3} m_\nu(0) e^{\xi \sigma_3}
\ ,
\ee
where $\kappa^\prime=\kappa + 2 \chi$, $\xi=\chi t$, and we have ignored the
$t$-dependence of the coupling constant so that 
$\int \kappa dt \simeq \kappa t$,
etc.

   It is convenient to factor out the determinant 
\be  m_\nu = \sqrt{m_1 m_2} M \ . \ee
Then ,
\bea
\sqrt{m_1(t) m_2(t)} & = & e^{-\kappa^\prime t} \sqrt{m_1(0) m_2(0)} \\
M(t) & = & e^{\xi \sigma_3} M(0) e^{\xi \sigma_3} .
\eea 
Thus, while the overall scale $\sqrt{m_1m_2}$ has a simple exponential 
dependence on $t$, the running of $M$, which contains the mass ratio and 
the mixing angle, is just a seesaw transformation defined in the previous
section. 
The difference from the traditional seesaw model is that, instead of
$\xi \gg 1$, for the RGE running $\xi$ is usually small ($\sim 10^{-5}$ in the
SM). Nevertheless, the exact and analytic formulae given in 
Eqs.(\ref{eq:tomega}-\ref{eq:ceta}) are valid solutions of the RGE. Detailed
analysis of their properties will  be given in a separate paper.
We only note that, according to Eq.(\ref{eq:inv}), there is a (complex)
RGE invariant,
\be
\sin 2 \theta(t) \sinh 2 \bar{\lambda}(t) = \sin 2 \theta(0) \sinh 2 \bar{\lambda}(0) \ ,
\ee
where $\theta$ and $e^{4\bar{\lambda}}$ ($\bar{\lambda}=\lambda+ i \phi$)
 are, respectively,
the physical neutrino mixing angle and the mass ratio. This equation can be 
used to determine $\lambda(t)$ and $\phi(t)$, once $\theta(t)$ is obtained
from Eq.(\ref{eq:ttheta}). We should also emphasize that, because the solution
can exhibit resonant behavior, large effect can result even for very small 
running ($\xi \ll 1$).

\subsection{Parametrization of the three flavor matrix}
\label{sec:3flsub}

	As is clear from the previous discussions, the Euler 
parametrization is the most convenient for dealing with the two flavor problem.
The generalization to three flavor, then, amounts to parametrizing an $SU(3)$
element in the form (phase)(rotation)(phase).  However, there are altogether
eight parameters in $SU(3)$ while each phase matrix
 can only accommodate two.  So
there must also be an additional phase matrix contained in the rotational
part of a general $SU(3)$ matrix.  This decomposition is of course none other
than the familiar CKM matrix decomposition.  Thus, for three flavors, 
the analog of Eq.~(\ref{eq:V}) is
\begin{equation} \label{eq:V3}
V_{R}=e^{i(\varepsilon_{3}\lambda_{3}+\varepsilon_{8}\lambda_{8})}
      e^{i\varepsilon_{7}\lambda_{7}}e^{i\varepsilon_{5}\lambda_{5}}
	e^{i\delta_{3}\lambda_{3}}e^{i\varepsilon_{2}\lambda_{2}}
       e^{i(\varepsilon'_{3}\lambda_{3}+\varepsilon'_{8}\lambda_{8})}
\end{equation}
Like the CKM representation, the phase factor $e^{i\delta_{3}\lambda_{3}}$
could be put in a different location, or one could use another diagonal
$\lambda$ matrix.

	The seesaw problem for three flavors again aims at rewriting the
matrices so that $m_{\nu}$ is given as in Eq.~(\ref{eq:mW}), with $W$
assuming the form of Eq.~(\ref{eq:V3}).  As in the two flavor case,
the exterior phase factors of $W$ do not contribute to neutrino 
oscillations.  An
exact solution for the three flavor problem, however, is not easily
obtained owing to the complexity of computing finite matrices involving 
the $\lambda$ matrices.  In Sec.~\ref{sec:3flav}, we will present an 
approximate solution to the three neutrino problem.

\section{A unified approach to fermion mass matrices}

  Our general analysis of the properties of the seesaw model suggests a
unified picture of the quark and neutrino mass matrices.
As was discussed in Sec.~\ref{sec:2flav}, the physical mixing angle 
of a seesaw model is quite small, in general, but can be maximal when
special conditions are met. We will now present arguments which can 
associate these regions to the quark and neutrino mass matrices, respectively.
For simplicity, our discussions are restricted to the case of two flavor 
only.

\subsection{Quark mass matrices}

  It has been known for a long time that quark mass matrices can be 
adequately described in a seesaw form\cite{quarkss,FN} 
\be \label{eq:qm}
m = \left(\ba{cc} \mu_1 & \\ & \mu_2 \ea\right)
    \left(\ba{cc} a & b \\ b & c \ea\right)
 \left(\ba{cc} \mu_1 & \\ & \mu_2 \ea\right) \ ,
\ee
where $a,b,c$ are arbitrary complex numbers, all of the same order, and
$\mu_2/\mu_1 \gg 1$.
Here we have used the arbitrariness in $m$  to demand that it be complex
and symmetric, in contrast to the usual choice that $m$ is hermitian.

  If the physical masses are denoted as $m_1$ and $m_2$, then Eq.(\ref{eq:qm})
implies that, for generic values of $a,b,c$, 
\be
m_1/m_2 \sim (\mu_1/\mu_2)^2 \ .
\ee
While the mixing angle satisfies the well-known relation\cite{quarkss}
\be  
\sin^2 \theta \simeq m_1/m_2 \ .
\ee
This result was derived first for real matrices, and remains valid for
complex case, as discussed in Sec.~\ref{sec:2flav}. It has served as a model
for quark mass matrices for a long time.

 Physically, a symmetric and complex mass matrix can be derived by a
symmetry argument. Since the mass  term in the lagrangian is given by 
$\bar{q}_L M q_R$, a symmetric mass matrix can be naturally obtained by 
imposing a discrete $Z_2$ symmetry:
\be 
\bar{q}_L \leftrightarrow  q_R \ .
\ee
If we further impose a gauged horizontal symmetry, such as a $U(1)$ symmetry
{\it a la} Froggatt and Nielsen \cite{FN}, 
then we are led to a mass matrix in the 
form of Eq.(\ref{eq:qm}). For instance, we may take the horizontal charge
assignments $(0,1)$ for $(q_{L1}, q_{L2})$ and $(1,0)$ for $(q_{R1}, q_{R2})$.
The charge assignments for the mass matrix $\bar{q}_L M q_R$ is
\be
Q_M \sim \left(\ba{cc} 2 & 1 \\ 1 & 0 \ea\right) \ .
\ee 
The Froggatt-Nielsen mechanism then calls  for a mass matrix of a form
\be
M \sim \left(\ba{cc} \epsilon^2 a & \epsilon b \\
                  \epsilon b & c \ea\right) \ ,
\ee
as in Eq.(\ref{eq:qm}).

\subsection{Neutrino mass matrices}
\label{sec:numm}

In Sec.~\ref{sec:ST}, we have found that, in order to have a large physical
mixing angle, it is necessary that the $(2,2)$ element of $M_R^{-1}$ be small,
{\it i.e.}, $|C|\simeq 0$ for 
$M_R^{-1} = \left(\ba{cc} A & B \\ B & C \ea\right)$.
Since we have used the normalization $det M_R^{-1} = +1$, the Majorana mass
matrix is given by
\be
M_R = \left(\ba{cc} C & -B \\ -B & A \ea\right) \ .
\ee
The condition $|C| \simeq 0$ simply means that $M_R$ is itself 
of the seesaw form.  The condition $|C| \simeq 0$ is not sufficient, however,
to guarantee a large mixing angle, which is a consequence of further
constraints on $M_R$. We will not attempt a detailed model construction
here. We only note that, as emphasized in Secs.~\ref{sec:2flav} and 
\ref{sec:GP}, the mixing angle is very sensitive to small variation of the
parameters in $M_R$.  In particular, if a model is based on symmetry 
arguments, symmetry breaking effects have to be weighed carefully.

 In summary, both the quark and neutrino mass matrices can be adequately 
described in the seesaw form. Their difference arises from the Majorana
sector, which is itself of the seesaw form. This last requirement can lead
to large mixing in the effective neutrino mass matrix. The sensitivity to
small changes in the parameters calls for a careful examination which should
also include three flavor effects. Detailed model construction along these
lines will be attempted in the future.

\section{An approximate solution to the three flavor problem}
\label{sec:3flav}

	In Sec.~\ref{sec:3flsub}, it was pointed out that the 
three flavor seesaw~\cite{3nu}
problem amounts to rearranging products of matrices in $SL(3,C)$.  
Since a general, analytical, solution is not available, we will turn to an
approximate solution which is physically relevant.

	For the three neutrino problem, it is known that the (23) angle is near
maximal, the (13) angle is small, and that the (12) angle is probably large.
This suggests that, to a good approximation, the three flavor problem can be
decomposed into two, two flavor problem.  To implement this scenario, let us
consider the $3 \times 3$ matrix $M_{R}^{-1}$,
\begin{equation}
M_{R}^{-1}=\left(\begin{array}{ccc}
                A & B & D \\
                B& C & E \\
               D& E & F \\
                   \end{array}
                  \right)
\end{equation}
The neutrino matrix, with $m_{D}$ diagonal and $U=I$ for simplicity of
presentation, since the general case can be easily incorporated as in
Eq.~(\ref{eq:MN}), is given by
\begin{equation} \label{eq:M3} 
m_{\nu}=\left(\begin{array}{ccc}
                m_{1} &  &  \\
                & m_{2} &  \\
               &  & m_{3} \\
                   \end{array}
                  \right)
\left(\begin{array}{ccc}
                A & B & D \\
                B& C & E \\
               D& E & F \\
                   \end{array}
                  \right)
\left(\begin{array}{ccc}
                m_{1} &  &  \\
                & m_{2} &  \\
               &  & m_{3} \\
                   \end{array}
                  \right).
\end{equation}

	It is convenient to introduce, in addition to the Gell-Mann
$\lambda$ matrices, $\lambda_{9}$ and $\lambda_{10}$,
\begin{equation}
	\lambda_{9}=\left(\begin{array}{ccc}
                0 &  &  \\
                & 1 &  \\
               &  & -1 \\
                   \end{array}
                  \right),
\end{equation}
\begin{equation}
\sqrt{3} \lambda_{10}=\left(\begin{array}{ccc}
                -2 &  &  \\
                & 1 &  \\
               &  & 1 \\
                   \end{array}
                  \right).
\end{equation}
We may now write
\begin{equation}
\left(\begin{array}{ccc}
                m_{1} &  &  \\
                & m_{2} &  \\
               &  & m_{3} \\
                   \end{array}
                  \right)=
\left(\begin{array}{ccc}
                m_{1} &  &  \\
                & \sqrt{m_{2}m_{3}} &  \\
               &  & \sqrt{m_{2}m_{3}} \\
                   \end{array}
                  \right)
e^{-\xi \lambda_{9}},
\end{equation}
\begin{equation}
\xi=\frac{1}{2}\ln(m_{3}/m_{2}).
\end{equation}
Then, the (23) submatrix of $m_{\nu}$ can be diagonalized,
\begin{equation}
e^{-\xi \lambda_{9}}
       \left(\begin{array}{ccc}
                A & B & D \\
                B& C & E \\
               D& E & F \\
                   \end{array}
                  \right)
e^{-\xi \lambda_{9}}=
U \left(\begin{array}{ccc}
                A & B' & D' \\
                B' & \Lambda' & 0 \\
               D' & 0 & \Sigma' \\
                   \end{array}
                  \right)U^{T},
\end{equation}
\begin{equation}
     U=e^{i\alpha \lambda_{9}}e^{i\beta \lambda_{7}}e^{i\gamma \lambda_{9}},
\end{equation}
\begin{equation}
   e^{-\xi \lambda_{9}}
       \left(\begin{array}{c}
                 A \\
                 B  \\
                D \\
                 \end{array}
                  \right)
             =U\left(\begin{array}{c}
                A \\
                        B'  \\
                D' \\
                 \end{array}
                  \right),
\end{equation}
and ($\Lambda',\Sigma'$) are the eigenvalues.
Although we could have chosen a proper normalizing factor so that the
(23) submatrix has $det=\pm 1$, as in Sec.~\ref{sec:2flav},
 for this problem it is
simpler not to do this and $\Lambda' \Sigma' \neq 1$, in general.
If we absorb $e^{i\gamma \lambda_{9}}$ by defining the new variables
($B'',D'',\Lambda'',\Sigma''$)=
($e^{i\gamma}B',e^{-i\gamma}D',e^{2i\gamma}\Lambda',e^{-2i\gamma}\Sigma'$),
and since $\lambda_{7}$ and $\lambda_{9}$ commute with the remaining Dirac
matrix, Eq.~(\ref{eq:M3}) becomes
\begin{equation}
m_{\nu}=X \left(\begin{array}{ccc}
                A & B'' & D'' \\
                B'' & \Lambda'' & 0 \\
               D'' & 0 & \Sigma'' \\
                   \end{array}
                  \right) X^{T},
\end{equation}
\begin{equation}
	X=e^{i\alpha \lambda_{9}} e^{i\beta \lambda_{7}}
        \left(\begin{array}{ccc}
                m_{1} &  &  \\
                & \sqrt{m_{2}m_{3}} &  \\
                &  & \sqrt{m_{2}m_{3}} \\
                   \end{array}
                  \right).
\end{equation}

	Now, the (13) rotation is controlled by $\left|D''/\Sigma'' \right|$.
However, we must first make sure that they have the same phase
(with the approximation $\left|m_{1}^{2} A \right| \ll
\left|m_{2}m_{3}\Sigma'' \right|)$.  To this
end let us multiply $m_{\nu}$ by $e^{i\omega \sqrt{3}\lambda_{10}}$ on
either side, and choose $\omega$ so that $e^{-\omega}D''$ 
and $e^{2i\omega}\Sigma''$ have the same phase, 
$\arg(e^{-i\omega} D'')=\arg(e^{2i\omega}\Sigma'')$.  In this case, we can
rotate away the (13) element of $m_{\nu}$ without changing its other elements
by assuming that the angle of rotation is small,
$\left|m_{1}D'' \right| \ll \left|\sqrt{m_{2}m_{3}} \Sigma''\right|$. 
We have (with $A''=A-D''^{2}/\Sigma''$), approximately,
\begin{equation}
m_{\nu}\simeq Y\left(\begin{array}{ccc}
                A'' & B'' & 0 \\
                B'' & \Lambda'' & 0 \\
               0 & 0 & \Sigma'' \\
                   \end{array}
                  \right)Y^{T},
\end{equation}
\begin{equation}
	Y=e^{i\alpha \lambda_{9}} 
        e^{-i\omega(\sqrt{3} \lambda_{10})} e^{i\beta \lambda_{7}} 
           e^{-i\psi \lambda_{5}}e^{i\omega(\sqrt{3} \lambda_{10})}
	\left(\begin{array}{ccc}
                m_{1} &  &  \\
                & \sqrt{m_{2}m_{3}} &  \\
                &  & \sqrt{m_{2}m_{3}} \\
                   \end{array}
                  \right),
\end{equation}
\begin{equation}
\tan \psi=(m_{1}e^{-i\omega}D'')/(\sqrt{m_{2}m_{3}}e^{2i\omega}\Sigma'')=real.
\end{equation}

	After this somewhat laborious route, we see that the diagonalization
of $m_{\nu}$ can be finally achieved by working solely in the (12) sector.
The crucial assumption for the success of this procedure is that
$\tan \psi \ll 1$.  Otherwise the (13) rotation $e^{-i\psi \lambda_{5}}$
will generate non-negligible elements all over the matrix $m_{\nu}$.
Although the exact condition for $\tan \psi \ll 1$ seems complicated,
in practice, as long as the elements $B$ and $D$ in $M_{R}^{-1}$ are
reasonably small, the approximation is valid.

	Fortunately, it is known that in reality the physical (13) rotation
angle is small.  This means that for any successful $m_{\nu}$, the
above approximation is appropriate.  In this case, the three neutrino problem is reduced
to two, two-flavor problem.  In particular, two popular scenarios, the
bimaximal or single maximal models, can be accommodated.

\section{Conclusion}

	Recent experimental data have revealed two striking features of the
intrinsic properties of the neutrinos.  One, as expected, they are very light.
Two, perhaps surprisingly, at least some of their mixing angles are large,
or even maximal.  The seesaw model provides a natural explanation of the
lightness.  However, the story of the mixing angles is more complicated.
In the seesaw model, the neutrino mixing matrix can be written as $UW$, where
$U$ comes from the left-handed rotation which diagonalizes the Dirac mass
matrix, and $W$, defined in Eq.~(\ref{eq:W}), is induced from the 
right-handed sector of the model.  For two flavors, the analytic solution
for $W$ shows that, when there is a mass hierarchy in $m_{D}$, the mixing
angle in $W$ is greatly suppressed for most of the available parameter space.
However, in a very small region, which we exhibited explicitly in 
Sec.~\ref{sec:2flav},
the mixing angle can be large.  In  addition, this region may be divided
roughly into two parts.  In one, characterized by $\gamma \approx \pi/4$,
the physical neutrino masses are nearly degenerate.  In the other, in which the
Majorana mass eigenvalues are hierarchical, the neutrino masses can be
either hierarchical or nearly degenerate.  This behavior of $W$ has 
interesting theoretical implications.

	Since the neutrino mixing matrix is given by $UW$, there are three
obvious possibilities which can lead to large mixing.  A) $U$ contains large
angles but $W \simeq I$; B) both $U$ and $W$ contribute appreciably and they
add up to form large mixing; C) $U \simeq I$ but the large angle is in $W$.
Corresponding to these possibilities we have three different physical
scenarios.  A) With $W \simeq I$, the physical neutrino masses are highly
hierarchical.  The burden for the model builders is to find a credible
theory which makes $U$ almost maximal naturally.  B) This scenario seems
the least likely to be implemented.  This is a ``just-so" solution whereby
the Dirac and Majorana sectors must conspire to make the resultant angle
large.  C) Here, $U \simeq I$ is quite reasonable from quark-lepton
symmetry, which leads naturally to $U \sim U_{CKM}$.  The challenge is
to find a mechanism whereby the parameters in the seesaw model lies
naturally in the narrow range for large mixing.

    In Sec.~\ref{sec:numm}, 
we have identified a necessary condition for large mixing,
namely, that the Majorana mass matrix is also of the seesaw type.
This result suggests a universal seesaw mechanism for both the quark and
neutrino mass matrices. The quarks can take advantage of the general solution,
resulting in small mixing and hierarchical masses.
For neutrinos, the existence of $M_R$ can then lead to large mixing angles.
More detailed studies are necessary to implement this scenario. 

  Our general analysis of symmetric and complex matrices also has an 
immediate application to the RGE running of the neutrino mass matrices.
Exact and analytic solutions of the RGE are found, in addition to a
(complex) RGE invariant which relates explicitly the running of the
mixing angle, the mass ratio and its phase.

	The analyses given above are for the case of two flavors.
However, in the approximation of a small (13) angle, we have found that
the three flavor problem is reduced to two, two flavor problems.
We thus do not expect qualitatively different physics for this case.

	In conclusion, the neutrino mixing matrix (masses and mixing angles)
implied by the seesaw model has rather intriguing properties.  To accommodate
large mixing angles, there are just a few limited options available.
These conditions should be helpful in the search of a viable neutrino
mass matrix.  We hope to return to this topic in the future.

\bigskip

 {\bf Acknowledgements:} \ 
We thank Jim Pantaleone for very helpful discussions and thank the 
referee for useful correspondence.
T. K. K. and G.-H. W. are supported in part by DOE grant No.
DE-FG02-91ER40681 and No. DE-FG03-96ER40969, respectively.


\begin{figure}[t]
\centerline{\epsfysize=20.cm\epsfbox{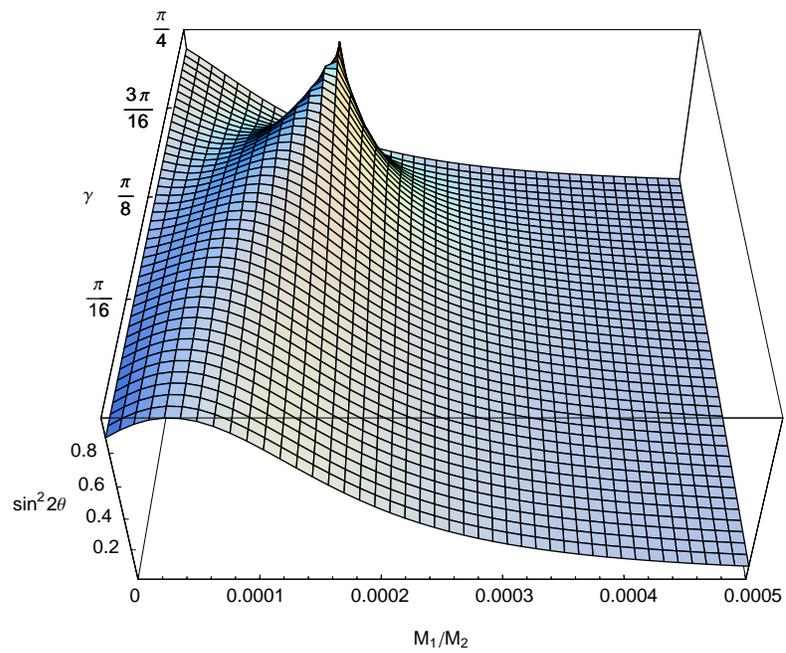}} 
\caption[] {A plot of $\sin^{2} 2\theta$ vs. ($M_{1}/M_{2}$) and $\gamma$,
with $\cos 2\beta = 0.9999$, $\tanh 2\xi = 0.9998$.  Note the
expanded scale of ($M_{1}/M_{2}$).  \label{fig2.ps}}
\end{figure}

\begin{figure}[t]
\centerline{\epsfysize=20.cm\epsfbox{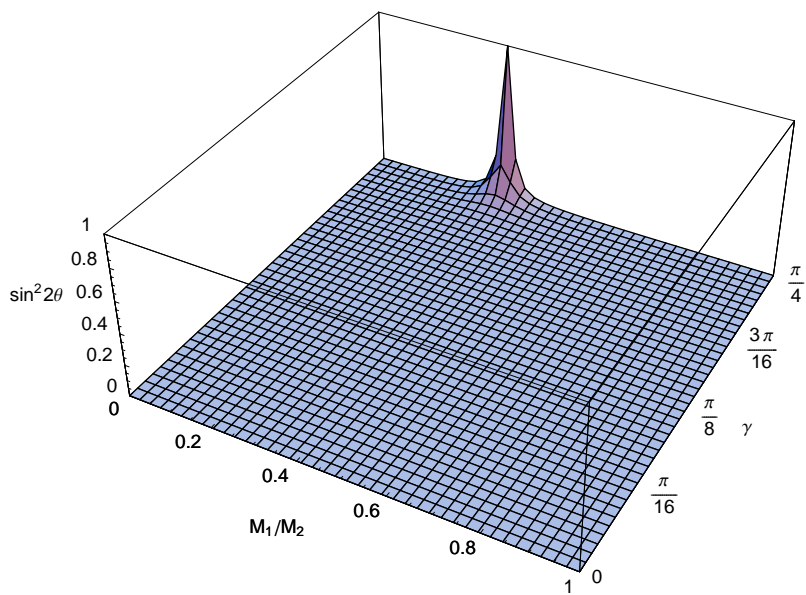}} 
\caption[] {Typical behavior of 
$\sin^{2} 2\theta$ for $\cos 2\beta < \tanh 2\xi$.  Here,
$\cos 2\beta = 0.5$, $\tanh 2\xi =0.9998$. \label{fig3.ps}}
\end{figure}

\begin{figure}[t]
\centerline{\epsfysize=20.cm\epsfbox{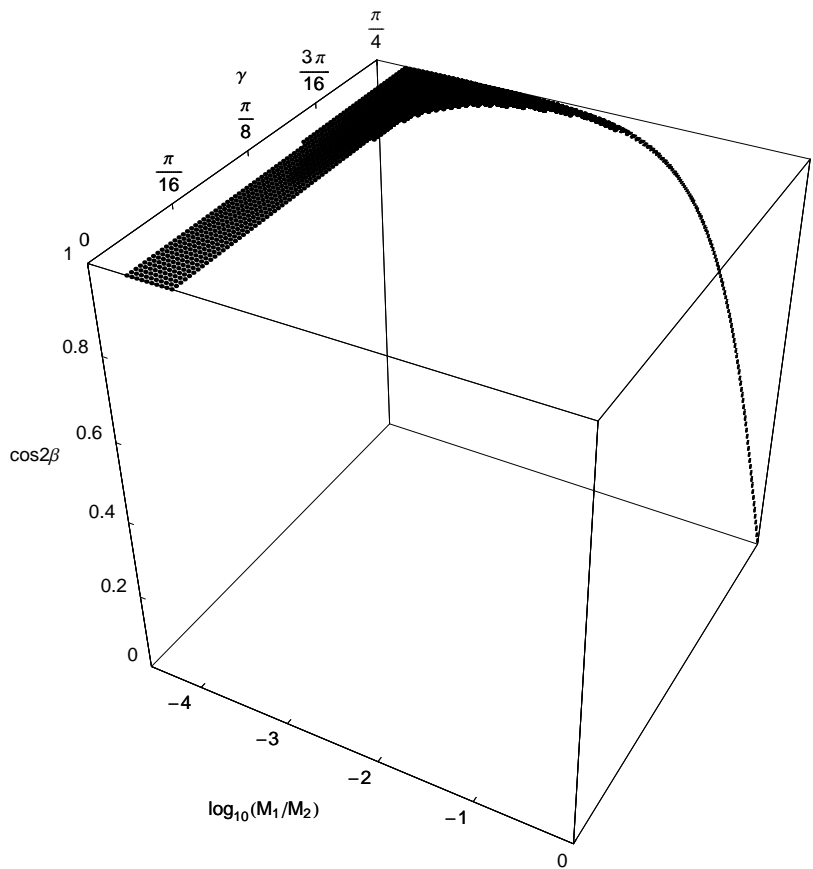}} 
\caption[] {Region in which the physical neutrino masses are nearly degenerate,
with $\mu_{1}/\mu_{2} > 0.5$, $\tanh 2\xi = 0.9998$. \label{fig4.ps}}
\end{figure}

\end{document}